\documentclass[twocolumn,showpacs,preprintnumbers,amsmath,amssymb,epsfig,widetext]{revtex4}

\usepackage{graphicx}
\usepackage{dcolumn}
\usepackage{bm}
\usepackage{epsfig}

\def\la{\mathrel{\mathpalette\fun <}}
\def\ga{\mathrel{\mathpalette\fun >}}
\def\fun#1#2{\lower3.6pt\vbox{\baselineskip0pt\lineskip.9pt
  \ialign{$\mathsurround=0pt#1\hfil##\hfil$\crcr#2\crcr\sim\crcr}}}

\input epsf

\newenvironment{tablehere}{\def\@captype{table}}{}
\newcommand{\tableskip}{\\[-6pt]}
\def\be{\begin{equation}}
\def\ee{\end{equation}}
\def\ba{\begin{eqnarray}}
\def\ea{\end{eqnarray}}

\def\nn{\nonumber}

\begin{document}

\preprint{}

\title{Large-Scale Tests of the DGP Model}

\author{Yong-Seon Song,$^{1,2}$ Ignacy Sawicki,$^{1,3}$  and Wayne Hu$^{1,2}$ }
\email{ysong@cfcp.uchicago.edu}
\affiliation{{}$^1$ Kavli Institute for Cosmological Physics, Enrico Fermi
Institute,  University of Chicago, Chicago IL 60637 \\
{}$^2$ Department of Astronomy \& Astrophysics,  University of Chicago, Chicago IL 60637\\
{}$^3$ Department of Physics,  University of Chicago, Chicago IL 60637
}

\date{\today}

\begin{abstract}
The self-accelerating braneworld model (DGP) can be tested from
measurements of the expansion history of the universe and the
formation of structure. Current constraints on the expansion history
from supernova luminosity distances, the CMB, and the Hubble
constant exclude the simplest flat DGP model at about 3$\sigma$. Even
including spatial curvature, 
best-fit open DGP model is a marginally poorer fit to
the data than flat $\Lambda$CDM.   Moreover, its substantially different
expansion history raises serious challenges for the model from
structure formation.
A dark-energy
model with the same expansion history would predict a highly
significant discrepancy with the  baryon oscillation measurement due
the high Hubble constant required. For
the DGP model to satisfy this constraint new non-linear phenomena
must correct this discrepancy.
Likewise
the large enhancement of CMB
anisotropies at the lowest multipoles due to the ISW effect would
require either a cut off in the initial power or new phenomena at the
cross over scale.    A prediction that is robust to both possibilities
 is that
high-redshift galaxies should be substantially correlated with the
CMB through the ISW effect.  This correlation should provide a sharp
test of the DGP model in the future.
\end{abstract}



\maketitle

\section{introduction}

Cosmic acceleration may arise either from a new form of energy
density, dubbed dark energy, or from a modification of gravity on cosmological scales.
Dvali, Gabadadze and Porrati (DGP)
\cite{dvali00} have proposed a braneworld modification of gravity to
explain acceleration (see \cite{lue05} for a recent review).
In this model our
universe is a (3+1)-dimensional brane embedded in an infinite
Minkowski bulk.  The
cosmological solution of this theory exhibits self-acceleration on
the brane classically
\cite{deffayet00,deffayet01}.

Various cosmological probes which can distinguish DGP from
dark energy models (DE) have been investigated in the literature
\cite{lue04,song04,schimd04,knox05,ishak05,amarzguioui05,linder05,sawicki05,koyama05,zhang05,stabenau06,maartens06}.  These come in two classes: those that test DGP modifications
of the
expansion history and those that test changes in how cosmological structures
are formed.
The advantage of the former class is that the predictions of the DGP model
are well understood theoretically.  The main disadvantage is that they currently lack
the power to distinguish the DGP model from its dark energy counterpart
the $\Lambda$CDM model at high significance.  Moreover, expansion rate tests
cannot distinguish between the DGP model and an arbitrary form of smooth dark
energy that is constructed to mimic it in the background.  Structure formation
tests on the other hand can in principle solve both these problems.  However
 due to complexities in the theory they currently can
only be rigorously implemented on a rather small range of scales.

In this paper, we examine the combination of these two types of
tests. We employ three tests of the expansion history: the angular
diameter distance to recombination from the three-year WMAP
data~\cite{spergel06}, luminosity distance to high-redshift
supernovae (SNGold\cite{riess04} and SNLS\cite{astier05}) and the
local expansion rate from Hubble constant measurements.  These tests
do in fact strongly disfavor the simplest version of the DGP model:
a modification of gravity in a spatially flat cosmology with no bulk
cosmological constant.   The best fit DGP model has a significant
negative curvature and a high Hubble constant.  Even with the
addition of spatial curvature, it is a slightly poorer fit to the
data than flat $\Lambda$CDM.

The best fit DGP model yields an expansion history that is strikingly different from
flat $\Lambda$CDM.   The dark energy model that mimics this expansion history
has negative curvature and a dark energy equation of state that increases sharply
with redshift.   In the dark energy context such an expansion history would
be strongly excluded by structure formation tests in weakly to strongly non-linear
regime.  These include
the baryon acoustic oscillations (BAO) in galaxy clustering \cite{eisenstein05,maartens06},
 the abundance
of galaxy clusters,  and fluctuations in the Lyman-$\alpha$ forest.   While these probes likely
also test the DGP model,
none of these phenomena are sufficiently well understood in the model to provide a definitive test. 
Indeed the presence of ghost degrees of freedom in the DGP model around
a de-Sitter background
\cite{Luty:2003vm,Nicolis:2004qq,Gorbunov:2005zk,Charmousis:2006pn}
require that the nature of gravitational interactions change
dramatically in the strongly-coupled non-linear regime for the model
to remain a viable explanation of cosmic acceleration \cite{Deffayet:2006wp}.  

Under this assumption, 
structure formation in DGP is currently well understood on scales between
 the Hubble scale and the scale radius of a typical dark matter
halo.  Large scale CMB anisotropy is substantially enhanced by the Integrated
Sachs-Wolfe (ISW) effect.   At the lowest multipoles, this excess poses
a significant challenge for the model.  Possible resolutions of this
problem still leave a robust prediction and sharp test of the DGP model: 
high redshift galaxies should be substantially correlated with the CMB.

\section{Expansion History}\vfill
\label{sec:tests}

\subsection{Supernovae, CMB and $H_{0}$}

The expansion history of a DGP model is determined by the usual
Friedman-Robertson-Walker parameters and the crossover distance.
The crossover distance is defined as the ratio of 5-dimensional to
4-dimensional Planck mass scales
\ba
r_\text{c}=\frac{M^{(4)\,2}_\text{Pl}}{2M^{(5)\,3}_\text{Pl}}.
\ea
It is a free parameter of the theory. If $r_\text{c}$ is close to
the current horizon scale, the acceleration of cosmic expansion is
replicated without dark energy.
Gravity on scales much less than $r_\text{c}$ does not penetrate
into the bulk significantly, and exhibits the usual Newtonian
potential falling as the inverse distance.
In the DGP model, as the Hubble scale
approaches $r_c$, the universe enters an accelerating,
asymptotically de-Sitter phase despite the absence of a cosmological
constant.

The main evidence for cosmic acceleration is geometrical and based
on the luminosity distance to high-redshift supernovae (SN)
\cite{riess98,perlmutter98,perlmutter99,riess04} and the angular
diameter distance to the recombination epoch of the CMB (e.g.
\cite{spergel03}).  Both distances are given as usual by the
(comoving) angular diameter distance in units of $H_0^{-1}$
\ba
 D(z)={1\over \sqrt{\Omega_k}} \sinh \left( r(z) \sqrt{\Omega_k}
\right),
\ea
where $r(z)$ is
\ba
 r(z)=\int^z_0 {dz'  \over E(z')}.
\ea
$E(z)$ is the Hubble parameter in units of the Hubble
constant.  The DGP model modifies the Friedman equation or the
relationship between $E(z)$ and the energy densities of matter
and radiation in units
of the critical density
\ba
E^2 &\equiv&  \frac{H^2}{H^2_0} \\
&=&\Omega_k a^{-2}+\left(
\sqrt{\Omega_\text{rc}}+\sqrt{\Omega_\text{rc}+\Omega_{
m}a^{-3} + \Omega_{r}a^{-4} } \right)^2\,, \nn
\ea
where $H_0=2997.9 h$ Mpc$^{-1}$.
Here the scale factor $a=1/(1+z)$,
 $\Omega_\text{rc}$(=$1/4r_\text{c}^2H_0^2$) is given by
\ba
\Omega_\text{rc}=\frac{(1-\Omega_k-\Omega_{m} - \Omega_{r})^2}{4(1-\Omega_k)}.
\ea
We implement the SN constraints from the ``gold"
SN data set of \cite{riess04}
and the SNLS set of \cite{astier05}.
For the CMB, we fix the distance to
recombination at $z_{\rm lss} = 1088^{+1}_{-2}$ through the measurement of the
acoustic peak scale ${\sc l}_A = 302^{+0.9}_{-1.4}$ and its length calibration
 through the matter density $\Omega_{m} h^2=0.1268
^{+0.0072}_{-0.0095}$ \cite{spergel03}.  Note that the SN are a
relative distance indicator and hence constrain $D$ whereas the CMB
is an absolute distance measure which constrains $D/h$.

\begin{figure}[htbp]
  \begin{center}
  \epsfysize=3.3truein
  \epsfxsize=3.3truein
    \epsffile{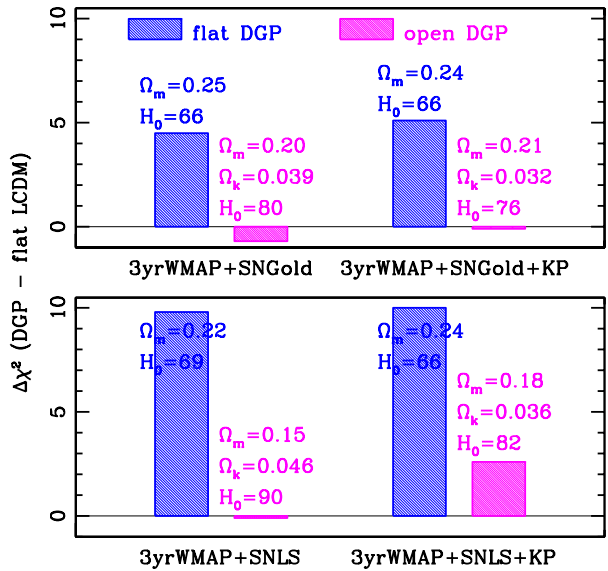}
    \caption{\footnotesize
The $\Delta\chi^2$ between the best fit
flat and open DGP versus that of flat $\Lambda$CDM model.
SNGold supernova (SN) data set is used
in the top panel and SNLS SN data set is used in the bottom panel.  The
DGP model requires curvature and a high Hubble constant.  With the
addition of Key Project (KP) direct Hubble constant measurements open
DGP is a marginally poorer fit to the data than flat $\Lambda$CDM.
}
\label{fig:dchi}
\end{center}
\end{figure}

Like flat $\Lambda$CDM (f$\Lambda$CDM), flat DGP has a single
parameter $\Omega_{m}$ that is over-constrained by the distances to
recombination, $\Omega_{m}h^{2}$, and high-redshift supernovae.
Unlike f$\Lambda$CDM, flat DGP is not a good fit to these data.
In DGP cosmologies, the SN favor a lower $\Omega_{m}$ than in
f$\Lambda$CDM resulting in a shorter angular diameter distance to the
CMB than is measured by WMAP.  For the SNGold data set the excess
$\chi^{2}$ over  f$\Lambda$CDM is $\Delta \chi^{2} \approx 4$
while for SNLS the excess is $\Delta\chi^{2}\approx 9$ (see
Fig.~\ref{fig:dchi}).

A negatively curved geometry elongates the CMB distance and allows a low
$\Omega_{m}$ in agreement with the SN distances
\cite{fairbairn05,maartens06}.  Even so, the
CMB constraint on $\Omega_{m}h^{2}$ favors a high Hubble
constant of $h= 0.8$ (SNGold) and $h\approx 0.9$ (SNLS)
 compared with observations from the HST Key Project (KP)
$h=0.72\pm 0.08$ \cite{KP}.
f$\Lambda$CDM on the other hand, despite being highly over-constrained
satisfies the observations with $h=0.73$.

With the addition of the KP $H_{0}$ constraint, the best fit
open DGP (oDGP) model has $h=0.76$ (SNGold) and $h=0.82$ (SNLS)
but is a marginally poorer fit to the data than f$\Lambda$CDM.
The excess over f$\Lambda$CDM of the latter is
 $\Delta\chi^2\sim 2.5$ with one fewer degree of freedom due
 to the fit to spatial
 curvature.
 The other parameters of this best fit
model are $\Omega_{m}=0.18$, $\Omega_{k}=0.036$.

  If the Hubble constant
measurements can be improved by a factor of 2 in the future, then the oDGP
and f$\Lambda$CDM will be clearly distinguished from each other.  They are
currently both acceptable fits to these data.
 We will hereafter take the best fit model to the WMAP, SNLS and
 KP data sets as the oDGP model for further tests.

 \subsection{DGP vs. Dark Energy}

The main expansion history differences between the best fit oDGP model and f$\Lambda$CDM
can be exposed by finding the scalar field or QCDM model that matches
the oDGP distance measures exactly.    With the same curvature, $\Omega_{m} h^2$ and $H_{0}$, that QCDM model has
a time-dependent equation of state
\ba
w(a)=-\frac{1}{3}\frac{\frac{dE^2}{d\ln a}
+3\Omega_ma^{-3}+2\Omega_ka^{-2}+4\Omega_ra^{-4}}
{{E}^2-\Omega_ma^{-3}-\Omega_ka^{-2}-\Omega_ra^{-4}}-1\,.
\ea
This equation of state starts at $w=-1/3$ during radiation domination,
becomes $w=-0.5$ during matter domination and decays to
$w=-0.85$ at the present time.
The value of $w(a)$ is substantially higher than the
$w=-1$ of f$\Lambda$CDM.
The implied increase in the dark energy density causes both the
larger Hubble constant and a decrement in
$D(z)$ at high redshift.  This decrement is compensated by negative
curvature to match the
CMB distance.

The expansion history of
this QCDM model is strikingly different from f$\Lambda$CDM despite
the fact that they both satisfy CMB and SN constraints.
Both the increased dark energy density and the negative spatial
curvature contribute to changes in structure formation that would
violate even current observations.    One therefore expects that
structure formation tests should provide a strong constraint on the
oDGP model.   It is important to emphasize that it is not necessary
to distinguish between oDGP and QCDM based on difference in the growth
of structure.  The comparison is between oDGP and f$\Lambda$CDM
because the QCDM model is not viable.
We therefore use the degenerate QCDM model as a guide to
understand the qualitative predictions of oDGP.
The main task of testing the expansion history consequences for
oDGP is to find a structure formation probe that can be robustly
calculated in the DGP model.

\section{Structure Formation}

\subsection{Baryon acoustic oscillations}

As discussed in the previous section, the largest expansion history
discrepancy between oDGP and f$\Lambda$CDM lies in the current
expansion rate or Hubble constant.   In the distance-degenerate QCDM
model, the large Hubble constant would be in strong conflict with
the measured baryon acoustic oscillations in the SDSS galaxy
correlations.  Baryon acoustic oscillations appear near the scale
where density perturbations become non-linear and hence cannot be
considered a purely geometric test. Detailed predictions in the oDGP
model below this scale require the solution of the $N$-body problem
with a non-linear Poisson equation.    It is nonetheless useful to
quantify the extent of the problem in the QCDM model since it
determines the level at which nonlinear oDGP corrections would have
to alter linear theory predictions to satisfy the data \cite{maartens06}.

The baryon oscillations are imprinted at a fixed physical scale
by the CMB at recombination. Under ordinary gravity, they yield
a nearly geometric test since non-linearities
are insufficient to change the physical scale of features substantially.
The SDSS luminous red galaxy (LRG) survey
 measures the features between
 $0.16\le z \le 0.47$.  At these redshifts, this distance is
 sensitive mainly to the Hubble constant \cite{EisHuTeg99a}.
 More specifically,
 it can be converted to a relative distance measure between
 the effective LRG redshift and recombination of
 $D(z=0.35)/D(z_{\rm lss})=0.0979\pm 0.0036$.  Recall that the
 CMB measures $D(z_{\rm lss})/h$.  Hence
 in the high Hubble constant QCDM model  this number is too low,
 specifically
 $0.0818$  leading to a 4.5$\sigma$ discrepancy.  This discrepancy can be
 slightly ameliorated by refitting the models to all data including
 the BAO data \cite{maartens06}.

 If oDGP shared the same non-linear effects with QCDM, the BAO
 measurement would exclude it as well.
 In the DGP model, there is a key distance scale, $r_*$, above which the
linearized perturbations experience the weak-brane phase
\cite{gruzinov01,porrati02,lue02,dvali02,lue05}. In the weak-brane
regime $r_c \gg r \gg r_*$, gravity is modified as a scalar-tensor
theory.  Below $r_*$, gravity is described by the standard Einstein theory.
For a point mass $M$, the transition scale
\ba
r_*=(r_\text{c}^2
r_\text{g})^{1/3}\,, \,\,\,\,(r_\text{g}=2G_\text{N}M).
\ea
Since even the linear density field is constructed from the spatial
average of collapsed dark matter halos under any hierarchical structure formation
model, the critical scale is associated with the $r_{*}$ of the dark matter
halos.

For definiteness let us take the other cosmological parameters as in the
best fit QCDM model: $\Omega_{m}h^2=0.122$, $\Omega_{b}h^2 = 0.0224$,
$\tau=0.11$  with
a power law spectrum of adiabatic fluctuations with
$n_s=0.958$, $\delta_{\zeta} = 4.5 \times 10^{-5}$ at $k_{\rm norm}=0.05$ Mpc$^{-1}$.
In this QCDM model, the LRG galaxy number densities correspond
to halos of
$M_* \approx  10^{13} h^{-1} M_\odot$ corresponding to $r_{*}\approx 3$ Mpc.
%
This scale is only marginally smaller than the baryon oscillation scale.
Furthermore,
without a determination of how dark matter halos form under DGP and whether
the observed galaxies indeed reside in $10^{13} h^{-1}M_{\odot}$ halos it remains unclear whether
weak-brane phase predictions can be applied to the BAO measurements.

\subsection{Linear Growth Rate}

The same expansion history discrepancy between QCDM and
f$\Lambda$CDM which raises $h$ also causes both an earlier decay of
the growth of structure and a larger total change in the growth to
$z=0$.   For example, in the QCDM model $\sigma_{8}=0.6$ and
$\Omega_m=0.18$.  These values are in conflict with the abundance of
local clusters.   Unfortunately, this QCDM prediction again relies
on the non-linear aspects of structure formation and hence cannot
yet be applied to oDGP. In order to test oDGP we must find a probe
on scales $r_{c}\gg r \gg r_{*}$.

In this regime, the linear theory predictions of DGP are now well
understood. In general, the equations of motion of linear
perturbations in DGP are not closed on the brane.  They require
knowledge of the gradient of metric perturbations into the bulk as
quantified by the so-called master variable \cite{deffayet02a}. This
in turn depends on boundary conditions in the bulk. Koyama and
Maartens \cite{koyama05} solved this system in the small scale limit
$k/aH\gg 1$ where the expansion rate can be ignored compared to the
master variable frequency $\omega\sim k$. In this quasi-static (QS)
limit, the Newtonian metric perturbations become
\ba
\frac{k^2}{a^2}\Phi_{-}&=&4\pi G\rho_m\Delta_m\,, \nn\\
\frac{k^2}{a^2}\Phi_{+}&=&- \frac{4\pi G}{3\beta} \rho_m\Delta_m \,,
\ea
where $\Phi_{-}= (\Phi-\Psi)/2$ and $\Phi_{+}=(\Phi+\Psi)/2$ and
\ba
\beta=1-2r_c H\left(1+\frac{\dot{H}}{3H^2} \right).
\ea
Here $\Delta_m$ is the comoving matter density perturbation and
for $k/aH \gg 1$ satisfies
\ba
\ddot{\Delta}_m+2H\dot{\Delta}_m=-\frac{k^2}{a^2}(\Phi_{+}-\Phi_{-}).
\ea
These relations agree with the earlier conclusions of Lue \cite{lue04} obtained in
a different manner.

\begin{figure}[htbp]
  \begin{center}
  \epsfysize=3.3truein
  \epsfxsize=3.3truein
    \epsffile{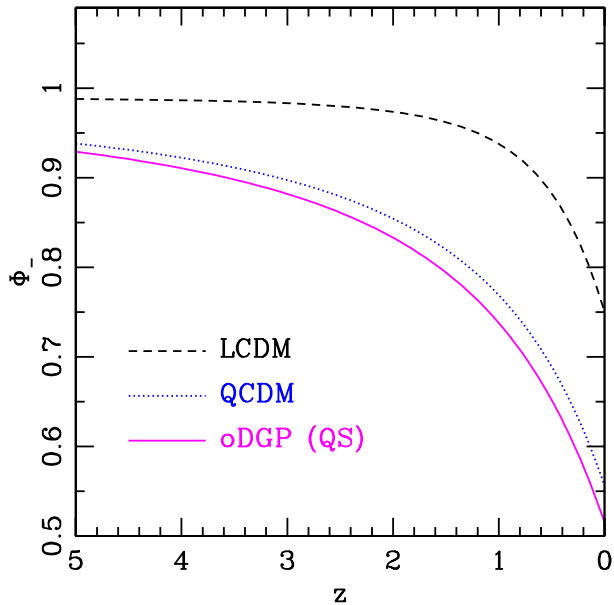}
    \caption{\footnotesize
The  growth factor of $\Phi_{-}$ of  f$\Lambda$CDM and oDGP under
the quasi-static (QS) assumption normalized to $\Phi_{-}(z\gg 1)=1$.
The model parameters are chosen as the
best fit to the CMB (3yrWMAP), SN (SNLS), and $H_{0}$ (KP). QCDM is the scalar
field dark energy model which has the same expansion history as oDGP.  The growth
rate is plotted on subhorizon scales where the dark energy is smooth and oDGP follows
the QS assumption.
}
\label{fig:Phi}
\end{center}
\end{figure}

This QS solution implies a 5$\%$ to 10$\%$ extra decay of
$\Phi_{-}$ in the  oDGP compared with the QCDM model with the
same expansion history.
We show the evolution of $\Phi_{-}$ for each model in Fig.~\ref{fig:Phi}.
The main effect however is not a difference between oDGP and QCDM but between either
and f$\Lambda$CDM.  In either case, $\Phi_{-}$ begins to decay at high-$z$
such that there is as much decay at $z>1$ as there is in f$\Lambda$CDM at $z<1$.

In order to find an appropriate test of this prediction, we need to
determine the validity of the quasistatic solution on large scales approaching
$r_{c}$.  In a companion paper \cite{sawicki06}, we solved the dynamical equations by
making a starting assumption that outside the horizon the evolution is scale
free and that perturbations in the bulk vanish at the causal horizon.  We then
iterate the solution through the dynamical equations until they reach convergence.
The starting scaling assumption is mildly violated during the transition from matter
domination to accelerated expansion but the iteration yields stable solutions.
We call this the dynamical scaling (DS) solution.

Under these assumptions the quasistatic solution is stably approached
by $k=10 aH$ and hence is valid to better than $2\%$ at
$k \ga 0.01$ Mpc$^{-1}$ for $z=0$.
Combined with the condition from $r_{*}$, a robust test of  the quasi-static predictions of
oDGP would
involve fluctuations between $0.01 \la k $(Mpc$^{-1}$) $\la 0.1$.

\subsection{ISW Effect}

The decay in the gravitational potential $\Phi_{-}$ at $k\ga 0.005$
Mpc$^{-1}$ causes an enhancement of the large-angle anisotropy in
the CMB through the ISW effect.

The angular power spectrum of the ISW effect is given by
\ba
C^{\rm II}_{\sc l}=4\pi \int \frac{{\rm d}k}{k}
[I^{\rm I}_{\sc l}(k)]^{2} { k^3 P_{\Phi_-\Phi_-}(k,0) \over 2\pi^2},
\label{eqn:ISWCl}
\ea
where $P_{\Phi_-\Phi_-}(k,0)$ is the power spectrum of
$\Phi_{-}$ at the present time and
the kernel $I^{\rm I}_{\sc l}$ is
\ba
I^{\rm I}_{\sc l}(k)=\int {\rm d}z \frac{\Phi_-(k,z)}{\Phi_-(k,0)} W^{\rm I}(k,z)
j_{\sc l}(kD) \left( {d D \over d r} \right)^{1/2}\,.
\ea
The window function is given by
\ba
W^{\rm I}(k,z)&=&
\frac{2}{\Phi_-}\frac{\partial\Phi_-}{\partial z}\,.
\ea
We have have here assumed that the spatial curvature is small enough that
the spherical Bessel function $j_{\sc l}$ accurately represents the radial harmonics.

\begin{figure}[htbp]
  \begin{center}
  \epsfysize=3.3truein
  \epsfxsize=3.3truein
    \epsffile{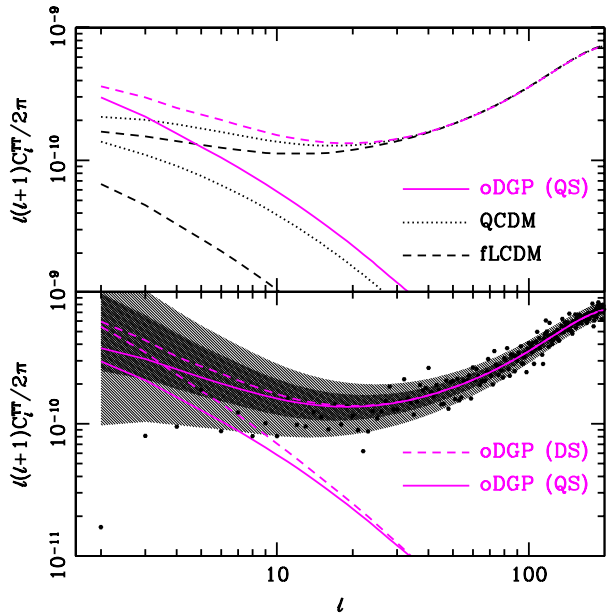}
    \caption{\footnotesize
    The CMB temperature anisotropy power spectra of the various models.
   Top panel: oDGP under the quasi-static (QS) approximation versus QCDM and
   f$\Lambda$CDM for the total power (upper curves) and the ISW effect
   (lower curves).  Bottom panel: oDGP (QS)
   compared with the dynamical scaling (DS) solution.  The DS solution predicts
   a further enhancement of the ISW effect such that ${\sc l} <10$ strongly violates
   the WMAP measurements (points).  The errors are mainly associated with
   cosmic variance at these multipoles and the $68\%$ and $95\%$ CL bands per ${\sc l}$
   have been attached to the DS model.
}
\label{fig:cl}
\end{center}
\end{figure}

Fig.~\ref{fig:cl} shows the difference between predictions for
$C_{\sc l}^{\rm II}$ from oDGP under the quasistatic approximation
compared with f$\Lambda$CDM and QCDM.   We also show the total
temperature power spectrum $C_{\sc l}^{\rm TT}$. The enhancement of
the low-multipole power spectrum directly reflects the high-redshift
decay of the gravitational potential in oDGP and QCDM. In QCDM the
effect at the lowest multipoles is smaller since a scalar field has
a Jeans scale at the horizon.  Beyond this scale, dark energy
perturbations slow the decay of the potential.

\begin{figure}[htbp]
  \begin{center}
  \epsfysize=3.3truein
  \epsfxsize=3.3truein
    \epsffile{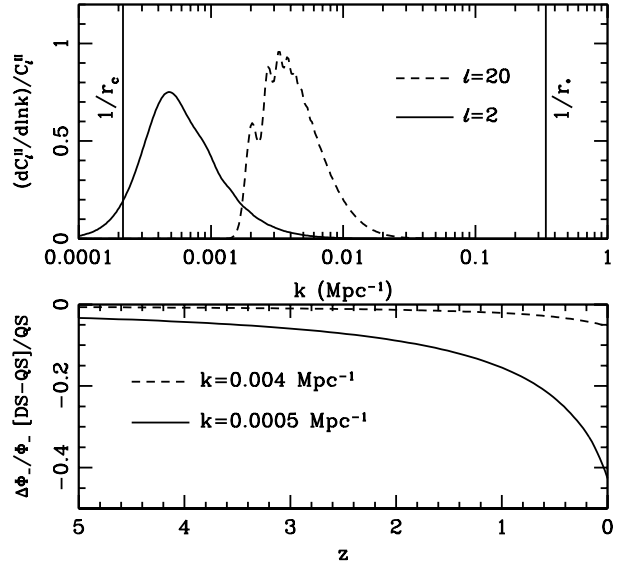}
    \caption{\footnotesize Top panel: the spatial wavenumbers corresponding to the ISW effect
    under the QS approximation
     for multipoles ${\sc l}=2, 20$.   For ${\sc l}=20$ the scales associated
    with the effect are $1/r_c \ll k \ll 1/r_*$ and even the quadrupole receives much of its
    contribution from $1/r_c \ll k$.  Bottom panel: the fractional difference between
    the  dynamical scaling (DS) and quasi-static (QS) solutions for DGP at the relevant scales
    for ${\sc l}=2,20$.  For ${\sc l}=20$, the QS approximation is accurate to the several percent
    level.
}
\label{fig:int_ISW}
\end{center}
\end{figure}

There are analogous corrections for oDGP on large scales but they
have the opposite sign.  Under the DS approximation the decay rate of
the $\Phi_{-}$ actually increases at the largest scales (see Fig. \ref{fig:int_ISW})
leading to even larger fluctuations.  In \cite{sawicki06}, we show that
this approximation should
at least give the right sign of the effect.  The QS solution should thus be taken
as a lower bound on the ISW effect.  Hence the discrepancy between
the oDGP QS predictions and the measurements at ${\sc l}=2-5$ in the 3year WMAP
data are a challenge to the model that will require either a cut off in
the initial spectrum or other new phenomena near $r_{c}$
to overcome.

At ${\sc l} \sim 20$ the ISW effect comes from physical scales that are within
the limit of the quasistatic approximation as well as  well below $r_{c}$.
The ISW effect here is a small
fraction of the total anisotropy and is best isolated through cross correlation
with tracers of the gravitational potential, e.g. galaxies.
Moreover the cross correlation can isolate the redshift
history of the potential decay and expose a key prediction of the oDGP
structure formation: an early decay of the potential.

\section{Galaxy-ISW correlation test}

\begin{figure*}[t]
\centerline{
\epsfxsize=5.3truein\epsffile{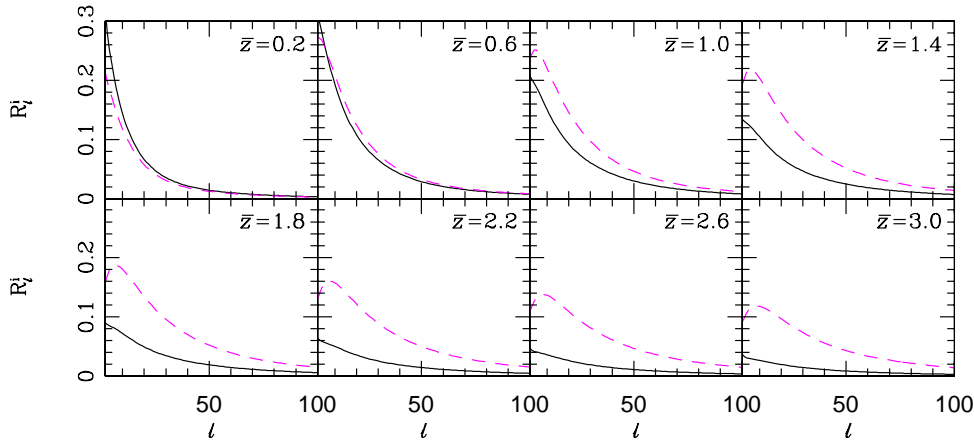}$\,$}
\caption{
The galaxy-ISW cross-correlation coefficient
$R^i_{\sc l}$ in each galaxy bin from $z=0$ to $z=3$.
Solid curves denotes f$\Lambda$CDM
and dash curves denotes oDGP.  Note the much larger
correlation at high $z$ in oDGP.
}
\label{fig:R}
\end{figure*}

The oDGP model predicts a stronger and earlier decay of the gravitational
potential than the f$\Lambda$CDM model.
By cross-correlating galaxies of different redshifts with the CMB,
one can in principle reconstruct the redshift history of the potential decay.
Furthermore, the cross-correlation arises from the well understood
quasi-static (QS) regime of oDGP.

The cross-power spectrum of the CMB and a set of galaxies $g_{i}$ is given by
a generalization
of Eq.~(\ref{eqn:ISWCl})
\ba
C^{g_{i}{\rm I}}_{\sc l}=4\pi \int \frac{{\rm d}k}{k}
I^{g_{i}}_{\sc l}(k)I^{\rm I}_{\sc l}(k){ k^3 P_{\Phi_-\Phi_-}(k,0) \over 2\pi^2},
\ea
where
the galaxy kernel $I^{g_{i}}$ is
\ba
I^{g_{i}}_{\sc l}(k)=\int {\rm d}z \frac{\Phi_-(k,z)}{\Phi_-(k,0)} W^{g_{i}}(k,z)
j_{\sc l}(kD) \left( {d D \over d r} \right)^{1/2}.
\ea
Under the QS approximation, the window function becomes
\ba
W^{g_{i}}(k,z)&=&\frac{2}{3\Omega_{m}}
\frac{k^2}{H_0^2}
\frac{n_{i}(z)b_{i}(z)}{1+z}\,,
\ea
where $n_{i}(z)$ is the redshift distribution
of the galaxies normalized to $\int dz n_{i} = 1$
and $b_{i}(z)$ is the galaxy bias.

For definiteness,
we assume that the galaxy sets come from a net galaxy distributions of
\ba
n_g(z)\propto z^2 e^{-(z/1.5)^2}\,,
\ea
where the normalization is given by the LSST expectation \footnote{{\tt http://www.lsst.org}}
of 35 galaxies per
$\rm arcmin^2$.
For the subsets of galaxies, we assume that this total distribution is
separated by photometric redshifts which have a Gaussian error
distribution with rms
$\sigma(z)=0.03(1+z)$ (see \cite{hu04} for details).
The redshift distributions are then given by
\ba
n_i(z)={A_{i}\over 2}n_g(z)
\left[ {\rm erfc}\left(\frac{z_{i-1}-z}{\sqrt{2}\sigma(z)}\right)
-{\rm erfc}\left(\frac{z_{i}-z}{\sqrt{2}\sigma(z)}\right)\right]\,,\nn
\ea
where erfc is the complementary error function and $A_{i}$ is determined
by the normalization constraint.

Our galaxy bias $b_{i}(z)$ is determined by a halo model and ranges from
$1.4$ at $z\sim 0$ to $3.8$ at $z \sim 3$.  However
with our narrow binning, the bias is nearly constant across the bin and
so its value can be empirically determined through comparing the
auto and cross power spectra once the data are in hand.
In particular,
the cross-correlation coefficient $R^i_{\sc l}$ in harmonic space
\ba
R^i_{\sc l}=\frac{C^{g_{i}{\rm I}}_{\sc l}}
{\sqrt{C^{{\rm TT}}_{\sc l}C^{{g_{i}g_{i}}}_{\sc l}}}
\ea
is independent of the bias and is a robust measure of the galaxy-ISW
correlation.

In Fig.~\ref{fig:R}, we show $R^i_{\sc l}$ for oDGP and
f$\Lambda$CDM. The main difference appears in the high-redshift bins
and reflects the early decay of the potential in oDGP.  The
correlation of high-redshift galaxies with the CMB is therefore a
sharp test of the oDGP model.

To determine whether these correlations are potentially observable,
we estimate the sampling
and noise errors
in the cross-correlation measurement.
First let us define the total galaxy power spectra
$\tilde{C}_{\sc l}^{g_{i}g_{i}}$
 as
\ba
\tilde{C}_{\sc l}^{g_{i}g_{i}}=C_{\sc l}^{g_{i}g_{i}}+N_{\sc l}^{g_{i}g_{i}}\,.
\ea
Here the shot noise $N_{\sc l}^{g_{i}g_{i}}=1/n^i_{A}$ where
$n^i_{A}$ is the angular number density of galaxies in the bin.
With our number densities the variance is nearly sample limited
in all of the redshift bins.

Likewise
the total power in the CMB measurements
is given by
\ba
\tilde{C}_{\sc l}^{\rm TT}=C_{\sc l}^{\rm TT}+N_{\sc l}^{\rm TT}
\ea
where the instrumental noise $N_{\sc l}^{\rm TT}$ is
$\Delta_{T}^2 e^{{\sc l}({\sc l}+1)\sigma^{2}/8{\rm ln}2}$.  With
either WMAP or
Planck satellite assumptions for the noise $\Delta_{T}$ and the
beam $\sigma$ the temperature map is sample variance limited to good approximation.

The error on cross-correlation can be calculated under the
Gaussian assumption as
\ba
\Delta C_{\sc l}^{g_{i}{\rm I}}=\sqrt{\frac{1}{(2{\sc l}+1)f_{\rm sky}}}
\left[ \left(C_{\sc l}^{g_{i}{\rm I}}\right)^{2}+
\tilde{C}_{\sc l}^{g_{i}g_{i}}\tilde{C}_{\sc l}^{\rm TT}\right]^{1/2}\,,
\ea
where $f_{\rm sky}$ is the sky fraction covered by the galaxy survey.

\begin{tablehere}
\begin{table}[hbt]\small
\begin{center}
{\sc Statistical Properties of Galaxy Correlations}\\
\begin{tabular}{c|c|c|c|c|c|c|c|c}
\tableskip\hline\hline \tableskip $\bar{z}$ & $0.2$ & $0.6$ &$1.0$ &$1.4$ &$1.8$ &$2.2$ &$2.6$ &$3.0$ \\
\tableskip\hline\tableskip
$(S^i/N^i)/\sqrt{f_{\rm sky}}$&2.5 & 5.0 & 5.9 & 6.0 & 5.7 & 5.3 & 4.9 & 4.5 \\
\tableskip\hline \tableskip ${\sc l}_{50\%}$ & 12 & 18 & 23 & 26 & 29 & 31 & 33 & 35 \\
\tableskip\hline \tableskip $R_{{\sc l}_{50\%}}^{\rm oDGP}/R_{{\sc l}_{50\%}}^{{\rm f}\Lambda{\rm CDM}}$ & 0.83 & 1.1 & 1.6 & 2.2 & 2.8 & 3.6 & 4.4 & 5.3 \\
\tableskip\hline
\end{tabular}\\[12pt]
\caption{Rows: mean redshift, total signal-to-nose of
correlation, multipole ${\sc l}$ below which 50\% of the $(S^{i}/N^{i})^{2}$ arises, ratio
of correlation between oDGP and f$\Lambda$CDM at that multipole.}
\label{tab:R/N}
\end{center}
\end{table}
\end{tablehere}

The signal to noise in each bin is given by
\ba
\left(\frac{S^i}{N^i}\right)^2 = \sum_{\sc l}
\left(\frac{C_{\sc l}^{ g_{i}{\rm I}}}{\Delta C_{\sc l}^{{g_{i}{\rm I}}}}\right)^2
\label{eqn:signaltonoise}
\ea
and these values are given in Tab.~\ref{tab:R/N}.
In the sample variance limit approached by our fiducial assumptions
\ba
\left(\frac{S^i}{N^i}\right)^2 \approx \sum_{\sc l} {( 2{\sc l} +1)} { f_{\rm sky}}
{ (R_{\sc l}^{i})^{2} \over 1 +  (R_{\sc l}^{i})^{2} } \,.
\ea
Even at
high redshift $z\sim 2-3$ the signal to noise allows a $15\%-20\%$
measurement of the correlation.  Such a measurement would be more
than sufficient to distinguish the factor of a few difference
between the oDGP and f$\Lambda$CDM models.

Most of the detectable signal comes from the regime where the QS
approximation is appropriate.  At low multipoles ${\sc l}<10$, the
cosmic variance of both the galaxy and temperature fields dominate
\cite{Afs04}.  By ${\sc l}=100$ the sample variance of the acoustic
peaks dominates. From the  signal to noise sum in
Eq.~({\ref{eqn:signaltonoise}), one can define ${\sc l}_{50\%}$ as
the multipole at which the $(S^{i}/N^{i})^{2}$ reaches half its
total value. This scale is also given in Tab.~\ref{tab:R/N} and lies
between ${\sc l}=20-40$ for most of the bins.  At this scale the
ratio of correlation coefficients in oDGP and f$\Lambda$CDM
approaches a factor of 6 at $z=3$.  High redshift galaxies  can thus
test the quasi-static oDGP predictions with high significance.

\section{Discussion}

The self-accelerating DGP braneworld model is challenged by expansion history
constraints from the CMB, supernovae and the Hubble constant.   Unlike
a model with a cosmological constant as the dark energy, the DGP model
requires spatial curvature to satisfy the CMB and supernovae constraint
and even so mildly exceeds bounds on the Hubble constant.

In a dark energy model, the DGP expansion history would be ruled out
by structure formation tests.   These are serious but perhaps not
insurmountable challenges for the model.
Among the most powerful are the baryon oscillations
and the large angle CMB anisotropy from the ISW effect.
In the DGP model, these two phenomena
lie close to the two critical scales: $r_*$ where gravity transitions from obeying
the usual Einstein equations to a scalar-tensor or weak brane phase and $r_c$
when gravity exits the weak brane phase and becomes fully 5 dimensional.
At the $r_{*}$ transition, the theoretical framework to calculate structure formation
is currently incomplete.  At the $r_{c}$ scale, changes in the initial power
spectrum or new gravitational physics may ameliorate the problem.

The enhancement of the ISW effect from the decay of the
gravitational potential extends to intermediate scales where the
theoretical calculations are robust. The main difference between the
DGP  and the flat cosmological constant expansion history is the
prediction that the gravitational begins its decay at high redshift.
A sharp test of the DGP scenario would be to isolate this decay at
high redshift by cross correlating the CMB with a high-redshift
galaxy population.  The cross correlation coefficient is also
relatively robust to changes in the initial amplitude of fluctuations.
We show that a galaxy survey that covers a
substantial fraction of the sky at $z>1$ can detect this correlation
at high significance.   Such surveys are currently being planned for
the measurement of high-redshift baryon oscillations.  Alternately,
radio sources, $X$-ray sources and quasars provide other
high-redshift populations for cross correlation.

A possible source of confusion for this test is the apparent number density
fluctuation induced by magnification from gravitational lensing by the same
structures associated with the ISW effect \cite{Hui06}.  Magnification both
lowers the number density for a fixed population behind a lens and also
raises it for a fixed flux limit by bringing faint galaxies into the sample.  This effect may be
separated from a true spatial correlation by examining different populations
of objects where the competition between these two effects differ and the
cross-correlation between high and low redshift galaxy populations.  A full
examination of this issue is beyond the scope of this paper but it is
likely that the galaxy-ISW cross correlation will become a sharp test
of the DGP braneworld acceleration model in the future.

\vspace{1.cm}

\noindent {\it Acknowledgments}: We thank Sean Carroll, Yu Gao, Nemanja Kaloper,
Arthur Lue,
and Xiaomin Wang for useful conversations.
This work was supported by the
U.S.~Dept. of Energy contract DE-FG02-90ER-40560. IS and WH are
additionally supported by the David and Lucile Packard Foundation.
This work was carried out at the KICP under NSF PHY-0114422.

\bibliography{bias}

\end{document}